\documentclass[twocolumn,pra,showpacs,preprintnumbers,superscriptaddress,amsmath,amssymb,10pt,aps]{revtex4}

\usepackage{mathptmx}
\usepackage{subfigure}
\usepackage{psfrag,graphicx}
\usepackage{dcolumn}
\usepackage{amsmath,amssymb}
\usepackage{bm}
\usepackage{color}
\usepackage{latexsym}
\usepackage{epstopdf}
\usepackage{color}
\usepackage[english]{babel}
\usepackage{latexsym}
\usepackage{psfrag,graphicx}
\usepackage{subfigure}
\usepackage{amsmath}
\usepackage{amssymb}
\usepackage{amsfonts}
\usepackage{bm}
\usepackage{natbib}
\usepackage{epstopdf}
\usepackage{appendix}

\definecolor{mygrey}{gray}{0.35}
\definecolor{myblue}{rgb}{0.2,0.2,0.8}
\definecolor{myzard}{cmyk}{0,0,0.05,0}
\definecolor{mywhite}{rgb}{1,1,1}
\definecolor{mywhite}{rgb}{1,1,1}
\definecolor{myred}{rgb}{1,0.,0.3}

\usepackage[colorlinks=true,citecolor=myblue,linkcolor=myred]{hyperref}

\def\ba{\begin{align}}
\def\enda{\end{align}}
\def\bi{\begin{itemize}}
\def\ei{\end{itemize}}

\def\be{\begin{equation}}
\def\ee{\end{equation}}
\def\bea{\begin{eqnarray}}
\def\eea{\end{eqnarray}}
\def\bse{\begin{subequations}}
\def\ese{\end{subequations}}

\def\f{\text{f}}

\begin{document}
\title{Adiabatic flux insertion and growing of Laughlin states of cavity Rydberg polaritons}
\author{Peter A. Ivanov}
\affiliation{Department of Physics and Research Center OPTIMAS, University of Kaiserslautern, Germany}
\affiliation{Department of Physics, St. Kliment Ohridski University of Sofia, James Bourchier 5 blvd, 1164 Sofia, Bulgaria}
\author{Fabian Letscher}
\affiliation{Department of Physics and Research Center OPTIMAS, University of Kaiserslautern, Germany}
\affiliation{Graduate School Materials Science in Mainz, Gottlieb-Daimler-Strasse 47, D-67663 Kaiserslautern, Germany}
\author{Jonathan Simon}
\affiliation{Department of Physics and James Franck Insitute, University of Chicago, Chicago, IL, USA}
\author{Michael Fleischhauer}
\affiliation{Department of Physics and Research Center OPTIMAS, University of Kaiserslautern, Germany}

\begin{abstract}
Recently, the creation of photonic Landau levels in a twisted cavity has been demonstrated in
Nature \textbf{534}, 671 (2016). Here we propose a scheme to adiabatically transfer flux quanta in multiples of $3\hbar$ simultaneously to all cavity photons by coupling the photons through flux-threaded cones present in such cavity setup. The flux transfer is achieved using external light fields with orbital angular momentum and a near-resonant dense atomic medium as mediator. Furthermore, coupling the cavity fields to a Rydberg state in a configuration supporting electromagnetically induced transparency, fractional quantum Hall states can be prepared. To this end a growing protocol is used consisting of a sequence of flux insertion and subsequent single-photon insertion steps. We discuss specifically the growing of the $\nu=1/2$ bosonic Laughlin state, where we first repeat the flux insertion twice creating a double quasi-hole excitation. Then, the hole is refilled using a coherent pump and the Rydberg blockade.
\end{abstract}

\date{\today}

\pacs{
}
\maketitle


\section{Introduction}

\begin{figure}
\includegraphics[width=0.45\textwidth]{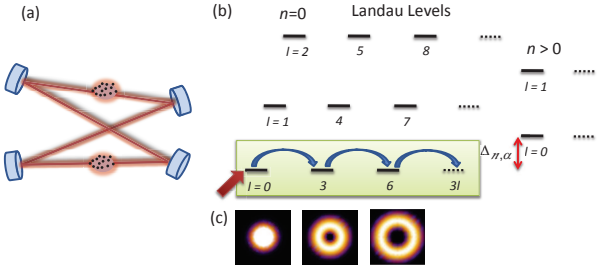}
\caption{(Color online) (a) Non-planar resonator consisting of four mirrors creating an artificial magnetic field for cavity photons.
Two clouds of atoms are used for adiabatic flux insertion and strong interactions mediated by Rydberg atoms.
(b) Structure of the photonic Landau levels labeled by orbital angular quantum number $l$ and radial quantum number $n$. Here $\Delta_{n,\alpha}$ is the frequency of the mode $(n,\alpha)$. The green box indicates the lowest Landau level (LLL). Initially, photons are pumped into the cavity mode with $n = l =0$ (red arrow). Then, adiabatic flux insertion transfers orbital angular momentum from a classical light beam to cavity photons increasing the total angular momentum (blue arrow). (c) The density plot of the first three cavity modes in the LLL with angular momentum $l=0$, $l=3$ and $l=6$.}
\label{fig1}
\end{figure}

Over the last few years, there has been a remarkable progress in the experimental realization and study of topological models for photons \cite{Khanikaev2013,Wang2009}. Prominent examples are the creation of topological band structures in systems of coupled optical waveguides \cite{Rechtsman2013,Hafezi2011,Jorg} and resonators \cite{Ningyuan2015}. Photonic systems offer a number of potential advantages for spatially and time-resolved manipulation and detection of topological states. The ability to create strong interactions by coupling photons to Rydberg states \cite{Peyronel2012,Gorshkov2011,Baur2014} offers furthermore the possibility to study many-body topological effects such as fractional quantum Hall physics or fractional Chern insulators \cite{Sorensen2005, Hafezi2007, Umucalilar2012, Carusotto2013}. To obtain sufficiently large interaction gaps, lattice-free systems are preferable. Since the presence of an effective magnetic field is equivalent to a Coriolis force in a plane perpendicular to the field, it can be created by rotation in continuous systems, as is used for ultra-cold neutral atoms \cite{Cooper1999, Wilkin2000, Shaeer2001, Schweikhard2004}. To apply this technique to photons, however, requires either to make use of the drag effect of a rotating dispersive medium \cite{Otterbach2010,Gotte2007} or to enforce an effective rotation using non-planar ring resonators that induce a rotation of light rays about the optical axis on each round trip, as is shown in Fig. \ref{fig1}(a). The latter technique has recently been experimentally implemented leading to the first demonstration of synthetic Landau levels (LL) for photons \cite{Schine2016,Sommer2016}. Photons in such optical cavities provide an excellent experimental platform for the realization of fractional quantum Hall physics due to the effective two-dimensional motion as well as the enhanced optical nonlinearity caused by mode confinement and cavity-enhanced interaction time \cite{Sommer2015}.

In this paper we discuss the creation of photonic Laughlin (LN)-type ground states in the setup of Ref. \cite{Schine2016}, using the
growing technique suggested in \cite{Grusdt2014,Letscher2015}. A key feature of the scheme is the controlled insertion of single photon and magnetic flux quanta into the cavity system. The latter creates a quasi-hole excitation in the center of the system and subsequently transfers the right amount of angular momentum. Then the hole is refilled by a coherent laser field creating a LN-type ground state. Besides being efficient our growing scheme has the advantage as compared e.g. to that suggested in \cite{Dutta2017} that quasi-holes created by photon decay are continuously pumped to the periphery of the LN-droplet, allowing to prepare a quantum-Hall liquid which is almost defect free in the center. The main challenge of the scheme is to insert an
integer amount of flux quanta. For this we propose an adiabatic method for transferring external orbital angular momentum (OAM) from classical light beams to the cavity photons by using light-matter interaction as a mediator. Specifically we consider the interaction between an ensemble of four-level atoms coupled to the cavity field and classical light and show that the adiabatic transfer of OAM to the cavity photons can be achieved by using stimulated Raman adiabatic passage (STIRAP) \cite{Vitanov2017} with Laguerre-Gauss laser beams.
The transfer is facilitated by an infinite set of cavity dark-state polaritons which are a superposition of light and collective matter excitations.

In order to realize a single photon coherent pump, the photonic cavity modes are coupled to a high lying Rydberg state of an atomic medium
under conditions of electromagnetically induced transparency \cite{Pritchard2010, Petrosyan2011, Peyronel2012}, which leads to a strong photon nonlinearity. Employing the resulting photon blockade, a single photon can be inserted into the system.
Repeating a sequence of magnetic flux and subsequent photon insertion leads to a growing of the liquid while keeping it in the LN ground state.

The paper is organized as follows: In Sec. \ref{Cavity_Structure} we introduce the mode structure of the non-planar cavity. In Section \ref{FInsertion} we discuss the flux insertion technique. In particular, we show that the magnetic flux quanta can be inserted into the system by using successive STIRAP processes. Further we discuss imperfections which limit the fidelity of our protocol. In Sec. \ref{SPI} we discuss the insertion of a single photon into the cavity by using the strong photon-photon interaction and the preparation of LN-type photon states using the growing technique. Finally, in Sec. \ref{C} we summarize our findings.

\section{Photon Cavity Setup}\label{Cavity_Structure}

We consider the resonator system realized in Ref.\cite{Schine2016} and shown in Fig. \ref{fig1}(a). The transversal confinement of photons by four curved mirrors of a ring resonator creates a two-dimensional gas of photons trapped in a harmonic potential with an effective mass. The non-planar geometry leads to a rotation of the transverse mode profile in a single roundtrip which is equivalent to the action of an effective magnetic field pointing in the direction of propagation plus an anti-binding centrifugal potential \cite{Sommer2016}. For sufficiently strong rotation the anti-binding potential can compensate the harmonic confinement and photonic Landau levels emerge. Such a configuration is however unstable and sensitive to astigmatism. Increasing the effective rotation even further eventually leads to another level configuration with large degeneracy, see Fig. \ref{fig1}(b). Here all cavity modes $f_{n,l}(r,\phi)$ with angular momentum $l=3m+\alpha$, where $m$ is integer value have the same frequency $\Delta_{n,\alpha}$ for a given radial quantum number $n$ and fixed value of $\alpha =0,1,2$. The transverse pattern of modes has a three-fold symmetry.
We use the term lowest Landau level (LLL) referring to the degenerate modes with radial quantum number $n=0$ and $\alpha=0$ (i.e. $l = 0,3,6,\ldots$).
The photonic Landau levels are described by the Hamiltonian
\begin{equation}
\label{eq:PhotonicLandauLevel}
\hat{H}_0 =\hbar \sum_{\alpha=0}^2\sum_{n,m=0}^{\infty} \Delta_{n,\alpha} \hat{a}_{n,3 m+\alpha}^{\dag}\hat{a}_{n,3 m+\alpha}.
\end{equation}
Here, $\hat{a}_{n,l}^{\dag}$ and $\hat{a}_{n,l}$ are the creation and annihilation operators of a cavity photon in mode $f_{n,l}$.
It is convenient to express the cavity field operator $\hat{\mathcal{E}}=\sum_{\alpha=0}^{2}\sum_n\hat{\mathcal{E}}_{n,\alpha}$ as
\begin{equation}
\hat{\mathcal{E}}_{n,\alpha}(r,\phi)=\sum_{m=0}^{\infty}f_{n,3m+\alpha}(r,\phi)\hat{a}_{n,3m+\alpha}.
\end{equation}
The cavity mode functions $f_{n,l}$ are described by the Laguerre-Gauss (LG) eigenmodes that carry orbital angular momentum $\hbar l$ in the transverse plane,
\begin{equation}
f_{n,l}(r,\phi)= C_{n,l}\,
x^{|l|}e^{il\phi}e^{-x^{2}}L_{n}^{|l|}
\left(2x^2\right).\label{basis}
\end{equation}
Here $x=r/w_0$, with $w_{0}$ being the cavity waist $C_{n,l}=
\sqrt{{2^{|l|+1}n!}/{\pi(|l|+n)!}}\, (1/{w_0})$,
and $L_{n}^{|l|}(x)$ are the LG polynomials. The mode functions $f_{n,l}$ fulfill the orthogonality relation
\begin{equation}
\int_{0}^{2\pi}\!\!d\phi \int_{0}^{\infty}\!\!dr\enspace r f_{n,l}^{*}(r,\phi)f_{n^{\prime},l^{\prime}}(r,\phi)=\delta_{l,l^{\prime}}\delta_{n,n^{\prime}}.
\end{equation}
We note that all modes except $l= 0$ have a vanishing amplitude at the origin $r=0$ and thus do not couple to atoms in the center of the
transverse mode profile.

\section{Flux Insertion without interaction}\label{FInsertion}

\subsection{Principle}

\begin{figure}
\includegraphics[width=0.45\textwidth]{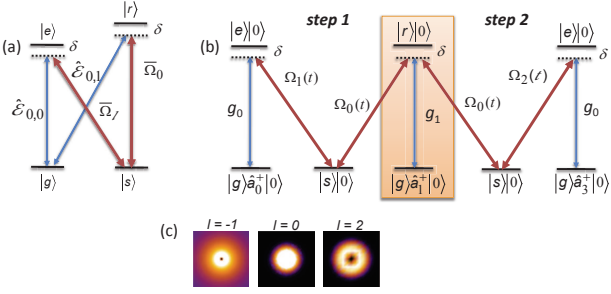}
\caption{(Color online) (a) The atomic level structure consists of two meta-stable levels $|g\rangle$, $|s\rangle$ and two excited levels $|e\rangle$ and $|r\rangle$. (b) System initially prepared in state with $l=0$. In the first stage of the protocol, the cavity modes $l=0$ and $l=1$ drives the transitions $|g\rangle\leftrightarrow|e\rangle$, $|g\rangle\leftrightarrow|r\rangle$ with coupling strength $g_{0}$ and $g_{1}$, while the laser fields $\Omega_{1}(t)$, $\Omega_{0}(t)$ couple the transitions $|e\rangle\leftrightarrow|s\rangle$, $|s\rangle\leftrightarrow|r\rangle$. We assume that the laser field $\Omega_{1}$ has an orbital angular momentum $-1\hbar$. As a consequence of this the initial photon in the $l=0$ mode is transferred to the $l=1$ mode using STIRAP technique. In the second step, the photon in the $l=1$ mode is transferred to the $l=3$ mode which belongs to the LLL manifold. This transition is performed by using the same atomic level structure, but with new Rabi frequency $\Omega_{2}$ which carriers OAM $2\hbar$. (c) Density plots of the classical driving fields. We assume that $\Omega_{0}$ Rabi frequency has a Gaussian shape with $l=0$. The other two classical laser fields carry OAM, and thus have vanishing intensity at the center.}
\label{fig22}
\end{figure}

The flux insertion protocol outlined below will lead to a controlled \emph{parallel} transfer of photons from modes $\hat{a}_{0,3m}$ to $\hat{a}_{0,3m+3}$ increasing the angular momentum per photon by $3\hbar$. To avoid any direct coupling between these two modes, which would lead to errors, we split the process into two successive steps:
\begin{equation}
{\rm i)}:\quad\hat{a}_{0,3m}\rightarrow \hat{a}_{0,3m+1},\qquad{\rm ii}): \quad \hat{a}_{0,3m+1}\rightarrow \hat{a}_{0,3m+3}.
\end{equation}
Light beams with OAM have already been successfully used to transfer angular momentum to an atomic medium \cite{Andersen2006}. Here we transfer orbital angular momentum from an external light beam to the cavity photons utilizing an atomic medium as a mediator. We consider atoms with four relevant states coupled  to the cavity field and external coherent driving fields carrying orbital angular momentum $l$ and with Rabi frequencies $\bar \Omega_{l}$ as indicated in Fig. \ref{fig22}(a).
The frequencies of all cavity modes with $n\ne 0$ (see Fig.\ref{fig1}(b)) are assumed to be far away from all atomic resonances and coupling to them is thus disregarded.
 In order to be able to switch on and off the coupling of the remaining cavity modes with $n=0$ to the atomic medium, we assume furthermore that also these transitions are sufficiently far away from single-photon resonance but that all Raman transitions are in two-photon resonance. In this way there is no interaction of the cavity field with the atomic medium in the absence of the classical driving fields.

   In the first step (i) two classical laser fields, $\bar\Omega_{-1}(r,\phi,t)$ and $\bar\Omega_{0}(r,t)$, which carry a net OAM of $1\hbar$ are applied to the atomic system. As a result a set of dark-state polaritons is created in the subspace of states with $n=0$ which are mixtures between the cavity field operators $\hat{a}_{0,3m}$ and $\hat{a}_{0,3m+1}$ and the corresponding matter components. By using time-varying laser fields in a STIRAP counterintuitive pulse order, photons are absorbed from modes $\hat{a}_{0,3m}$ and successively created in modes $\hat{a}_{0,3m+1}$. In this step an OAM of $1\hbar$ is transferred to each cavity photon in parallel.

   In the second step (ii) classical light field $\bar\Omega_{0}(r,t)$ and $\bar\Omega_{2}(r,\phi,t)$ with net OAM of $2\hbar$ are used which leads to the formation of a new set of dark-state polaritons now involving mode operators $\hat{a}_{0,3m+1}$ and $\hat{a}_{0,3m+3}$, see Fig. \ref{fig22}(b). In this step adiabatic following of the dark-state polariton transfers OAM of $2\hbar$ to each cavity photon. By successively repeating the processes one can increase the angular momentum of all occupied cavity modes in the lowest Landau levels in multiples of $3\hbar$.

\subsection{Atom-Field Interaction}

Consider an ensemble of atoms with a four-level atomic structure interacting with the cavity field as well as with an external classical light with OAM, Fig. \ref{fig22}(a). Here the transitions $\left|g\right\rangle\leftrightarrow\left|e\right\rangle$ and $\left|g\right\rangle\leftrightarrow\left|r\right\rangle$ are coupled to the cavity fields $\hat{\mathcal{E}}_{0,0}$ and $\hat{\mathcal{E}}_{0,1}$, while the atomic transitions $\left|s\right\rangle\leftrightarrow\left|e\right\rangle$ and $\left|s\right\rangle\leftrightarrow\left|r\right\rangle$ are driven by  classical light fields with time-dependent Rabi frequencies $\bar\Omega_{l}(r,\phi,t)$ and $\bar\Omega_{0}(r,t)$. The coupling Hamiltonian is given by
\begin{eqnarray}
&&\hat{H}=\hat{H}_{0}+\hat{H}_{\delta}+\hat{H}_{\rm c},\quad \hat{H}_{\delta}=\hbar\delta\int d^{2}r(\sigma_{ee}+\sigma_{rr}),\notag\\
&&\hat{H}_{\rm c}=-\hbar\int d^{2}r\Bigl[\sum_{m=0}^{\infty}g_{3m}f_{0,3m}(r,\phi)\hat{a}_{3m}\sigma_{eg}+\bar\Omega_{l}\sigma_{es}
\notag\\
&&\quad+\bar\Omega_{0}\sigma_{sr}+\sum_{m=0}^{\infty} g_{3m+1}f_{0,3m+1}(r,\phi)\hat{a}_{3m+1}\sigma_{rg}+{\rm h.c.}\Bigr].\enspace\label{H}
\end{eqnarray}
Here and in the following we have dropped the radial index $n$ of the cavity modes, since only the lowest Landau modes with $n=0$ are relevant.
The coupling strength $g_l$ are given by the atomic transition dipole matrix elements $d_{eg}$ and $d_{rg}$ of the ${\cal E}_{0,0}$ and ${\cal E}_{0,1}$ transitions, respectively, and overlap integrals with the mode functions
\begin{eqnarray}
&&g_{3m}  \sim  d_{eg}
\int_{0}^{2\pi}\!\!d\phi \int_{0}^{\infty}\!\!dr\enspace r f_{n,3m}(r,\phi)\, n(r),\notag\\
&&g_{3m+1}  \sim  d_{er}
\int_{0}^{2\pi}\!\!d\phi \int_{0}^{\infty}\!\!dr\enspace r f_{n,3m+1}(r,\phi)\, n(r),
\end{eqnarray}
where $n(r)$ is the atomic density.
As mentioned above, all transitions are assumed to be away from single-photon resonance with detuning $\delta$ but in respective two-photon resonance, such that turning off the classical light fields amounts to switching off the interaction of the cavity modes
with the atoms altogether. We assume $\Delta_{n\ne 0,\alpha}$ to be so large that the respective coupling can be disregarded. In Eq. (\ref{H}) we have introduced the standard continuous atomic flip operators $\sigma_{\mu,\nu}(\vec{r},t)=\frac{1}{\Delta V}\sum_{j\in\Delta V}|\mu_{j}\rangle\langle\nu_{j}|$ defined on a small volume $\Delta V$ centered around position $\vec{r}$ containing $\Delta N\gg 1$ atoms which fulfill the commutation relations $[\sigma_{\alpha,\beta}(\vec{r}),\sigma_{\mu,\nu}(\vec{r}^{\prime})]=\delta(\vec{r}-\vec{r}^{\prime})(\delta_{\beta,\mu}\sigma_{\alpha,\nu}(\vec{r})
-\delta_{\alpha,\nu}\sigma_{\mu,\beta}(\vec{r}))$.
In the following we are only interested in the weak probe regime meaning that to lowest order atoms remain in their ground state, therefore $\sigma_{gg}^{(0)}\approx 1$ \cite{Fleischhauer2005}. Thus, the only relevant operators for our discussion are the coherence of excited and ground states, $\hat{P}=\sigma_{ge}$ and $\hat{R}=\sigma_{gr}$ and the coherence between the two ground states $\hat{S}=\sigma_{gs}$. In the weak probe regime, these operators fulfill the commutation relation $[\hat{A}(\vec{r}),\hat{A}^{\dag}(\vec{r}^{\prime})]=\delta(\vec{r}-\vec{r}^{\prime})$ for $\hat{A}=\hat{P},\hat{R},\hat{S}$.
It is convenient to decompose them also into the LG basis (\ref{basis})
\begin{eqnarray*}
\hat{A}(r,\phi) &=& \sum_{n,l=0}^{\infty}\hat{A}_{n,l}\,  f_{n,l}(r,\phi).
 \end{eqnarray*}
In the following we discuss in details the two steps of the flux insertion protocol.

\subsection{First step}

In the first step of the flux insertion scheme we assume that the laser field $\bar\Omega_{l}(r,\phi,t)$ in Eq. (\ref{H})  has $l=1$, i.e.
carries an orbital angular momentum $-1\hbar$ such that
\begin{equation}
\bar\Omega_{1}(r,\phi,t)=\Omega_{1}(t)\, \kappa(x)\, e^{-i\phi},
\end{equation}
 with  $x=r/w_{0}$. The spatial profile $\kappa(x)$ will be chosen such that transitions to collective atomic states with $n>0$ are highly suppressed as we will show below. The latter is essential for our flux insertion scheme leading to a closed set of coupled states with different angular momentum $l$ but radial quantum number $n=0$ only.

The coupled system of Heisenberg-Langevin equations in the linear response regime are given by
\begin{eqnarray}
&&\frac{d}{d t}\hat{P}_{n,3m}=-(i\delta+\gamma)\hat{P}_{n,3m}+i\Omega_1\sum_{n^{\prime}=0}^{\infty}\chi_{3m}^{n,n^{\prime}}\hat{S}_{n^{\prime},3m+1}\notag\\
&&\quad\quad\quad\quad+i g_{3m}\hat{a}_{3m}\delta_{n,0}+\hat{F}_{n,3m}^{(p)},\notag\\
&&\frac{d}{d t}\hat{S}_{n,3m+1}= i\Omega_1^*\sum_{n^{\prime}=0}^{\infty}\left(\chi_{3m}^{n^{\prime},n}\right)^*\hat{P}_{n^{\prime},3m}+i\Omega_{0}\hat{R}_{n,3m+1},\notag\\
&&\frac{d}{d t}\hat{R}_{n,3m+1}= -(i\delta+\gamma)\hat{R}_{n,3m+1}+i\Omega_{0}^{*}\hat{S}_{n,3m+1}\notag\\
&&\quad\quad\quad\quad+i g_{3m+1}a_{3m+1}\delta_{n,0}+\hat{F}_{n,3m+1}^{(r)},\notag\\
&&\frac{d}{d t}\hat{a}_{3m}=ig_{3m}\hat{P}_{0,3m},\notag\\
&&\frac{d}{d t}\hat{a}_{3m+1}=ig_{3m+1}\hat{R}_{0,3m+1},\label{CS1}
\end{eqnarray}
where we use the orthogonality of the LG modes and have again dropped the radial index for the cavity modes. Here  $\gamma$ is the spontaneous decay rate from the excited states $|e\rangle$ and $|r\rangle$ which we assume for simplicity to be equal. $\hat{F}_{n,l}^{(p)}$ and $\hat{F}_{n,l}^{(r)}$ denote the respective Langevin noise operators which are delta-correlated in time and space. In the linear response regime, the population of the excited states is negligible, which allows to neglect the Langevin noise terms. Note that we also disregarded the cavity decay in our discussion. They will be discussed however in Sec \ref{SPI}. The matrix elements of the couplings between the states with radial quantum numbers $n$ and $n^{\prime}$ are given by
\begin{eqnarray*}
\chi_{3m}^{n,n^{\prime}}
&=& \int_{0}^{2\pi} d\phi \int_{0}^{\infty}dr \, r \kappa(x) e^{-i\phi} \, f_{n,3m}^*(r,\phi)\, f_{n^\prime,3m+1}(r,\phi).\\
\end{eqnarray*}
%

One recognizes from the coupled system (\ref{CS1}) that there is in general a coupling between modes with different radial index $n$.
We now choose the spatial profile of $\bar{\Omega}_1$ in such a way that couplings from the $n=0$ spin modes $\hat S_{0,3m+1}$ to higher modes with $n^\prime >0$ are highly suppressed, i.e. that $\chi_{3m}^{n,0} \sim \delta_{0,n}$.
This can be achieved if
\begin{equation}
\kappa(x) e^{-i\phi} f_{0,3m+1}(r,\phi) \sim f_{0,3m}(r,\phi),
\end{equation}
which would imply $\kappa(x)\sim 1/r$. This choice of the spatial profile is however not possible since $\bar\Omega_1$ carries a nonvanishing
OAM and thus must vanish for $r\to 0$. Thus we choose
\begin{equation}
\kappa(x) = \frac{x^2}{a^3+x^3},
\end{equation}
with $a=r_0/w_0$ and $r_0\ll w_0$ is some cut-off length.
We note that there is no choice of spatial profile that simultaneously perfectly suppresses the coupling of the $n=0$ optical polarization modes $\hat P_{0,3m}$ to modes with $n^\prime >0$. We will see however, that this is not necessary.

Let us first discuss the limiting case $a\to 0$. Then the coupling coefficient between spin and optical polarization within the LLL reads
\begin{equation}
\chi_{3m}^{0,0}=\sqrt{\frac{2}{3m+1}}.
\end{equation}
At the same time, while the couplings $\chi_{3m}^{0,n^\prime}$ between the LLL ($n=0$) with higher LL ($n^\prime \ge 1$) are all of order unity
\begin{equation}
\chi_{3m}^{0,n^\prime} = {\cal O}(1),
\end{equation}
the couplings
\begin{equation}
\vert \chi_{3m}^{n^\prime,0}\vert  \sim a^2 \ll \chi_{3m}^{0,0},
\end{equation}
are strongly suppressed in the limit $a\to 0$
(see Appendix \ref{LGP}).
In the following we therefore neglect the latter couplings but will discuss their effect later on.
Then the system of coupled equations simplifies to
\begin{eqnarray}
&&\frac{d}{d t}\hat{P}_{0,3m}=-(i\delta+\gamma)\hat{P}_{0,3m}
+i\sqrt{\frac{2}{3m+1}}\Omega_{1}\hat{S}_{0,3m+1}\notag\\
&&\quad\quad\quad\quad+i\Omega_1\sum_{n^\prime >0}^{\infty}\chi_{3m}^{0,n^\prime} \hat S_{n^\prime,3m+1}+i g_{3m}\hat{a}_{3m},\notag\\
&&\frac{d}{d t}\hat{S}_{0,3m+1}= i\sqrt{\frac{2}{3m+1}}\Omega_{1}\hat{P}_{0,3m}+i\Omega_{0}\hat{R}_{0,3m+1},\notag\\
&&\frac{d}{d t}\hat{R}_{0,3m+1}= -(i\delta+\gamma)\hat{R}_{0,3m+1}+i\Omega_{0}^{*}\hat{S}_{0,3m+1}+i g_{3m+1}\hat{a}_{3m+1},\notag\\
&&\frac{d}{d t}\hat{a}_{3m}=ig_{3m}\hat{P}_{0,3m},\notag\\
&&\frac{d}{d t}\hat{a}_{3m+1}=ig_{3m+1}\hat{R}_{0,3m+1}.\label{CS1234}
\end{eqnarray}
It is straightforward to show that from the coupled system (\ref{CS1234}) one can construct a dark-state polariton which is a constant of motion in the adiabatic limit
\begin{eqnarray}
\hat{\Psi}_{m}^{(1)}&=&\frac{1}{N_{m}}\Biggl\{g_{3m+1}\sqrt{\frac{2}{3m+1}}\Omega_{1}\,\hat{a}_{3m}+g_{3m}\Omega_{0}\hat{a}_{3m+1}\notag\\
&&\qquad - g_{3m}g_{3m+1}\hat{S}_{0,3m+1}\Biggr\},\label{dark1}
\end{eqnarray}
where $N_{m}(t)$ is the normalization factor.

Eq. (\ref{dark1}) shows that the light-matter interaction leads to the creation of an infinite set of dark-state polaritons labeled by the index $m$ which are a superposition of cavity-field operators $\hat{a}_{3m}$, and $\hat{a}_{3m+1}$ and corresponding collective ground-state coherences. Moreover, the dark-state polaritons have no contribution from the excited atomic states and thus are naturally immune to spontaneous decay.

Now a fully adiabatic transfer of excitations can be performed using a STIRAP protocol.
As long as $\Omega_{1}(t_{i})\gg\Omega_{0}(t_{i})\sqrt{\frac{3m+1}{2}}\frac{g_{3m}}{g_{3m+1}},g_{3m}$ the dark-state polaritons coincide with the initial state $\hat{\Psi}_{0,m}^{(1)}(t_{i})\simeq\hat{a}_{0,3m}$. Adiabatic following transfers the dark-state polaritons into $\hat{\Psi}_{0,m}^{(1)}(t_{1})\simeq\hat{a}_{0,3m+1}$ if $\Omega_{0}(t_{1})\gg\Omega_{1}(t_{\rm 1})\sqrt{\frac{2}{3m+1}}\frac{g_{3m+1}}{g_{3m}},g_{3m+1}$ which concludes the first step of the protocol. Thus at time $t_{1}$ the population from all modes $\hat{a}_{0,3m}$ of the LLL is transferred in parallel to modes $\hat{a}_{0,3m+1}$ which belong to an excited Landau manifold with $\alpha=1$.

So far we have discussed the limit $a\to 0$. Taking into account a small, but finite value of $a$ leads to some small residual couplings to
collective atomic modes with higher
radial index $n^\prime >0$. Since for those modes no dark state exists this coupling leads to some losses, which will be discussed in Sec.
\ref{Imperfections}.

In order to return the population back to the LLL manifold we repeat the same procedure as above using the same atomic structure but with new Rabi frequency $\bar\Omega_{2}(t)$ as we explain in the following.

\subsection{Second step}

The goal of the second step is to insert $2$ flux quanta and thus to return the photon to the LLL manifold. In order to perform this we assume that the transition $|e\rangle\leftrightarrow|s\rangle$ is driven with Rabi frequency
\begin{equation}
\bar\Omega_{2}(r,\phi,t)=\Omega_{2}(t)\tilde \kappa(x) e^{2i\phi},\label{Rabi22}
\end{equation}
 which carriers orbital angular momentum $2\hbar$. Consequently, the coupled system is similar
to Eq. (\ref{CS1})
\begin{eqnarray}
&&\frac{d}{d t}\hat{P}_{n,3m+3}=-(i\delta+\gamma)\hat{P}_{n,3m+3}+i\Omega_2 \sum_{n^{\prime}=0}^{\infty}\widetilde\chi_{3m+3}^{n,n^{\prime}}\hat{S}_{n^{\prime},3m+1}\notag\\
&&\quad\quad\quad\quad+i g_{3m+3}\hat{a}_{3m+3}\delta_{n,0}+\hat{F}_{n,3m+3}^{(p)},\notag\\
&&\frac{d}{d t}\hat{S}_{n,3m+1}= i\Omega_2^*\sum_{n^{\prime}=0}^{\infty}\left(\widetilde\chi_{3m+3}^{n^{\prime},n}\right)^*\hat{P}_{n^{\prime},3m+3}+i\Omega_{0}\hat{R}_{n,3m+1},\notag\\
&&\frac{d}{d t}\hat{R}_{n,3m+1}= -(i\delta+\gamma)\hat{R}_{n,3m+1}+i\Omega_{0}^{*}\hat{S}_{n,3m+1}\notag\\
&&\quad\quad\quad\quad+i g_{3m+1}a_{3m+1}\delta_{n,0}+\hat{F}_{n,3m+1}^{(r)},\notag\\
&&\frac{d}{d t}\hat{a}_{3m+3}=ig_{3m+3}\hat{P}_{0,3m+3},\notag\\
&&\frac{d}{d t}\hat{a}_{3m+1}=ig_{3m+1}\hat{R}_{0,3m+1},\label{CS123}
\end{eqnarray}
with new coupling matrix elements,
\begin{equation}
\widetilde{\chi}_{3m}^{n,n^{\prime}}= \int_{0}^{2\pi} d\phi \int_{0}^{\infty}dr \, r \tilde\kappa(x) e^{2i\phi} \, f_{n,3m}^*(r,\phi)\, f_{n^\prime,3m-2}(r,\phi).\\
\end{equation}

We now choose the spatial profile of $\widetilde\kappa(x)$ in such a way that couplings from the $n=0$ spin modes $\hat S_{0,3m+1}$ to higher modes with $n^\prime >0$ are highly suppressed, i.e. that $\widetilde\chi_{3m+3}^{n,0} \sim \delta_{0,n}$.
This can be achieved if
\begin{equation}
\widetilde\kappa(x) e^{2i\phi} f_{0,3m-2}(r,\phi) \sim f_{0,3m}(r,\phi).
\end{equation}
Thus we can simply set
\begin{equation}
\widetilde\kappa(x) = x^2.
\end{equation}
Note that with this choice all couplings of spin coherences $\hat S_{0,3m+1}$  to higher
states with $n^{\prime}>0$ are exactly cancelled
such that there are no undesired residual couplings. One has
\begin{equation}
\widetilde{\chi}_{3m+3}^{n^{\prime},0} = \frac{1}{2} \sqrt{\frac{(3m+3)!}{(3m+1)!}}\, \delta_{0,n^\prime},
\end{equation}
and the coupled system for the second stage of the protocol becomes
\begin{eqnarray}
&&\frac{d}{d t}\hat{P}_{0,3m+3}=-(i\delta+\gamma)\hat{P}_{0,3m+3}+i\Omega_2\widetilde\chi_{3m+3}^{0,0}\hat{S}_{0,3m+1},\notag\\
&&\quad\quad\quad\quad+i\Omega_2 \sum_{n^\prime >0}^{\infty} \widetilde\chi_{3m+3}^{0,n^\prime}\hat{S}_{n^\prime,3m+1}
 +i g_{3m+3}\hat{a}_{3m+3},\notag\\
&&\frac{d}{d t}\hat{S}_{0,3m+1}= i\Omega_2^*\, \widetilde{\chi}_{3m+3}^{0,0}\hat{P}_{0,3m+3}+i\Omega_{0}\hat{R}_{0,3m+1},\notag\\
&&\frac{d}{d t}\hat{R}_{0,3m+1}=-(i\delta+\gamma)\hat{R}_{0,3m+1}+i\Omega_{0}^{*}\hat{S}_{0,3m+1}+i g_{3m+1}\hat{a}_{3m+1},\notag\\
&&\frac{d}{d t}\hat{a}_{3m+3}= ig_{3m+3}\hat{P}_{0,3m+3},\notag\\
&&\frac{d}{d t}\hat{a}_{3m+1}=ig_{3m+1}\hat{R}_{0,3m+1}.\label{CS1235}
\end{eqnarray}
%
The adiabatic eigensolution of the coupled system (\ref{CS1235}) now are given by
\begin{eqnarray}
\hat{\Psi}^{(2)}_{m}&=&\frac{1}{\tilde N_{m}}\Biggl\{g_{3m+1}\frac{\Omega_{2}}{2}\sqrt{\frac{(3m+3)!}{(3m+1)!}}\hat{a}_{3m+3}
+g_{3m+3}\Omega_{0}\hat{a}_{3m+1}
\notag\\
&&\qquad -g_{3m+3}g_{3m+1}\hat{S}_{0,3m+1}\Biggr\}.
\label{dark2}
\end{eqnarray}
At time $t_{1}$ which is now the initial time for the second step the dark polariton (\ref{dark2}) coincides with  $\hat{\Psi}_{0,m}^{(2)}(t_{1})\simeq\hat{a}_{3m+1}$ as long as $\Omega_{0}(t_{1})\gg\Omega_{2}(t_{1})\frac{1}{2}\sqrt{\frac{(3m+3)!}{(3m+1)!}}\frac{g_{3m+1}}{g_{3m+3}},g_{3m+1}$. Adiabatically increasing $\Omega_{2}(t)$ drives the system into $\hat{\Psi}_{0,m}^{(2)}(t_{f})\simeq\hat{a}_{3m+3}$ if $\Omega_{2}(t_{f})\gg2\sqrt{\frac{(3m+1)!}{(3m+3)!}}\Omega_{0}(t_{f})\frac{g_{3m+3}}{g_{3m+1}},g_{3m+3}$ which concludes the second step. In total, the flux insertion protocol transfers orbital angular momentum in multiples of $3\hbar$ to all cavity modes of the LLL in parallel.
\begin{figure}
\includegraphics[width=0.45\textwidth]{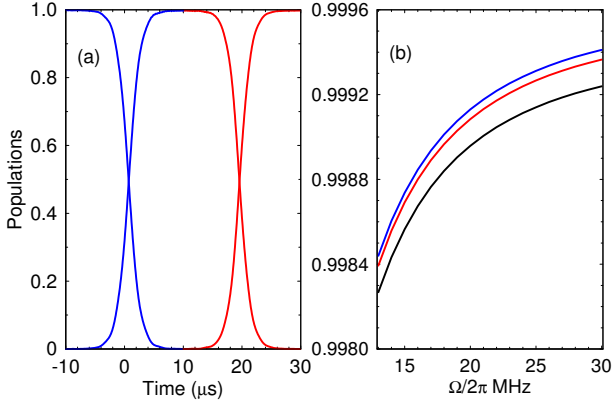}
\caption{(Color online) (a) The expectation values of $\hat{a}_{0,l}$ ($l=0,1,3$) versus time. The exact result for $\hat{a}_{0,0}$ and $\hat{a}_{0,1}$ (blue lines) according the coupled system Eq. (\ref{CS1}) including the residual couplings for $a=10^{-2}$. The red lines show the time-evolution of $\hat{a}_{0,1}$ and $\hat{a}_{0,3}$ according the coupled system Eq. (\ref{CS123}). We choose a time-dependent Rabi frequencies (\ref{Rabi}) and (\ref{Rabi2}) with $\Omega=2\pi\times 12.4$ MHz and cavity couplings $g=2\pi\times0.45$ MHz. The other parameters are set to $\gamma=0$, $\delta=2\pi\times 0.13$ MHz and $T=1$ $\mu$s. (b) The expectation value of $\hat{a}_{0,1}$ at time $t_{1}$ against $\Omega$ for $a=0.01$ (blue line), $a=0.015$ (red line), $a=0.02$ (black line).}
\label{stirap}
\end{figure}

\subsection{Imperfections}\label{Imperfections}
In the following we discuss the corrections that limit the fidelity of the adiabatic flux insertion. There are two main sources of imperfections. The first one are the residual off resonant couplings $\chi_{3m}^{n^\prime,0}$
to cavity modes with $n^{\prime}>0$. The second one is the violation of adiabaticity, in particular in the center of the atomic cloud, due to the vanishing amplitude of the classical light fields $\bar{\Omega}_{l}$ with $l\neq 0$.

\subsubsection{Residual couplings to higher collective atomic levels with $n>0$}
At the first stage of the scheme the residual coupling $\hat{S}_{0,3m+1} \leftrightarrow \hat{P}_{n,3m}$ with $n\ne 0$
 which we neglected in Eq. (\ref{CS1234}) leads to excitations of state $\left|e\right\rangle$ which spoils the adiabatic transition. We find that in the lowest-order this coupling strengths scale as
 \begin{equation}
\chi_{0}^{n^{\prime},0}\simeq -\frac{8\pi}{3}\sqrt{\frac{2}{3}} a^2,
\end{equation}
 for $m=0$ and become even smaller for higher $m$. Thus the condition $a\ll 1$ ensures the suppression of the undesired couplings to
 excited LL.

 As an example let us  consider the following time-dependent Rabi frequencies
\begin{equation}
\bar\Omega_{1}(r,\tau)=\kappa(x)\frac{\Omega}{\sqrt{1+e^{\tau}}},\quad \bar\Omega_{0}(\tau)=\frac{\Omega}{\sqrt{1+e^{-\tau}}},
\label{Rabi}
\end{equation}
which can be used to drive the first stage of the scheme. Here $\Omega$ is the peak-Rabi frequency and $\tau=t/T$ where $T$ is the characteristic pulse length. Assume that the pulses are applied in counterintuitive order in the time interval $[-\tau_{1},\tau_{1}]$ such that $\Omega_{1}(-\tau_{1})\gg\Omega_{0}(-\tau_{1})$ and respectively $\Omega_{1}(\tau_{1})\ll\Omega_{0}(\tau_{1})$. For the second stage we use
\begin{equation}
\bar\Omega_{2}(r,\tau)=x^2\frac{\Omega}{\sqrt{1+e^{2\tau_{1}-\tau}}},\quad \bar\Omega_{0}(\tau)=\frac{\Omega}{\sqrt{1+e^{\tau-2\tau_{1}}}},
\label{Rabi2}
\end{equation}
which guarantees that $\Omega_{2}(\tau_{f})\gg\Omega_{0}(\tau_{f})$. In Fig. \ref{stirap}(a) we plot the exact time evolution of the probabilities for adiabatic transition. As can be seen, including small cut-off length $a$ the respective
residual couplings do not affect the adiabatic flux insertion. Increasing the peak Rabi frequency $\Omega$ further suppresses the non-adiabatic transition, see Fig. \ref{stirap}(b).
\subsubsection{Non-adiabatic losses}

\begin{figure}
\includegraphics[width=0.45\textwidth]{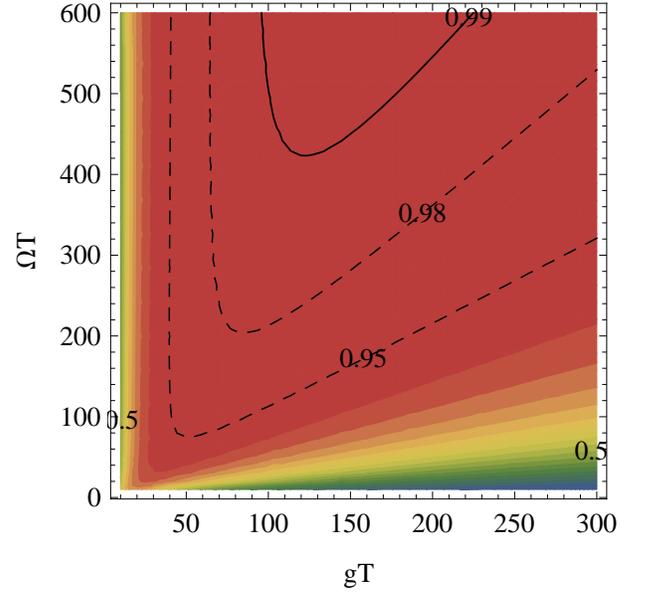}
\caption{(Color online) Fidelity of the flux insertion scheme versus the peak Rabi frequency $\Omega T$ and cavity coupling $g T$ according Eq. (\ref{fidelity}). We assume Gaussian distribution of the atomic density. The parameters are $\gamma T=100$, and $a=0.005$ and $\xi=0.25w_{0}$.    }
\label{fid}
\end{figure}

The orbital angular momentum transferred to the cavity photons in the flux insertion is taken from the classical laser fields $\bar{\Omega}_{l}$ with $l\neq 0$. These fields have necessarily a vanishing intensity at the origin $r=0$ and thus adiabaticity is violated in the center of the atom cloud. Instead of following adiabatically the dark state polariton, atoms close to the center will be excited into the intermediate states due to absorbtion of cavity photons. As a consequence of that there is a leak of population from the dark state subspace during both steps of the protocol. In the following we assume a worst-case scenario where the optical fields are on resonance. The finite detuning $\delta$ further suppress errors.

The total amount of non-adiabatic losses can be obtained by calculating the number of atoms not returning to the initial state after a full
cycle of the protocol.
The probability of an atom at distance $r$ from the center to remain in the adiabatic state
during  the first stage of the protocol can be estimated as, see Appendix \ref{NAL}
\begin{equation}
e_1(\vec{r})=\exp\biggl\{-\frac{2\gamma}{g^{2}}\int_{t_{i}}^{t_{1}} dt \Bigl(\dot{\varphi}_{1}^{2}\sin^{2}(\theta_{1})+
\dot\theta_{1}^{2}\cos^{2}(\theta_{1})\Bigr)\biggr\},\label{e1}
\end{equation}
where  we have assumed for simplicity that $g_{3m}=g_{3m+1}=g$. Here $\varphi_{1}(r,t)=\tan^{-1}\left(\bar\Omega_{0}(t)/\bar\Omega_{1}(r,t)\right)$ and $\theta_{1}(r,t)=\tan^{-1}\left(\sqrt{\bar\Omega_{1}^{2}(r,t)+\bar\Omega_{0}^{2}(r,t)}/g \right)$ are the time-dependent mixing angles. For the Rabi frequencies (\ref{Rabi}) one can evaluate the probability (\ref{e1}), see Appendix \ref{NAL}.

Similarly, the probability for an atom at position $r$ to remain in the dark state subspace during the second stage is given by
\begin{equation}
e_2(\vec{r}) =\exp\biggl\{-\frac{2\gamma}{g^{2}}\int_{t_{1}}^{t_{f}}\left(\dot{\varphi_{2}}^{2}\sin^{2}(\theta_{2})+
\dot{\theta_{2}}^{2}\cos^{2}(\theta_{2})\right)\biggr\},
\end{equation}
where now $\varphi_{2}(r,t)=\tan^{-1}\left({\bar\Omega_{2}(r,t)}/{\bar\Omega_{0}(t)}\right)$ and $\theta_{2}(r,t)=\tan^{-1}\left(
\sqrt{\bar\Omega_{2}^{2}(r,t)+\bar\Omega_{0}^{2}(t)}/g\right)$.

The total probability for all atoms to stay in the dark state after a full cycle of operation thus reads
\begin{equation}
p=\frac{\int d^{2}r\, \, n(\vec{r})\, e_1(\vec{r}) \, e_2(\vec{r})
}{\int  d^{2}r\, \, n(\vec{r})}.\label{fidel}
\end{equation}

Furthermore one has to take into account that the
system is prepared at the beginning of a full cycle in a state with a single photon in mode $f_{0,0}$ and all atoms in the ground state
$\vert g\rangle$. This state does not have a perfect overlap with the dark state, but the latter reads
\begin{equation}
p_{\textrm{in}}=\frac{1}{N}\int d^2r\, \frac{\Omega^{2}\kappa^2(r)}{g^{2}+\Omega^{2}\kappa^2(r)} \,n(\vec{r}),
\end{equation}
where we have used that initially $\bar\Omega_0(t_i)=0$.

This gives for the fidelity of the flux insertion
\begin{equation}
F= p\cdot p_\textrm{in}.\label{fidelity}
\end{equation}
In Fig. \ref{fid} we have plotted the fidelity as a function of the peak Rabi frequency $\Omega T$ and the cavity coupling $g T$ assuming Gaussian distribution of the density of atoms, $n(r)=n_{0}e^{-r^{2}/\xi^{2}}$ with $n_{0}=N/\pi\xi^{2}$. As can be expected increasing $g T$ compared to $\gamma T$ leads to smaller absorption length which improves the fidelity. However, for sufficiently high $gT$ compared to $\Omega T$ the respective overlap between the initial state and the dark state becomes smaller which limits the fidelity.

\section{Laughlin state preparation}\label{SPI}

Now, we discuss the preparation of Laughlin-type states in a setup of cavity Rydberg polaritons. Following Refs. \cite{Grusdt2014,Letscher2015}, a Laughlin state can be grown by the successive repetition of adiabatic flux insertion (see Sec. \ref{FInsertion}) and a single-photon coherent pump, discussed below.

\subsection{Rydberg Cavity Polaritons and Laughlin State}

To realize a fractional quantum Hall system requires besides the artifical magnetic field, discussed in Sec. \ref{Cavity_Structure} and Ref. \cite{Schine2016}, strong interactions between the photonic cavity modes in the lowest photonic Landau level,
\begin{equation}
\label{eq:PhotonPhotonInteraction}
\hat{H}_{\rm int}=\sum_{l_{1},l_{2}}\sum_{l_{3},l_{4}}V_{l_{3},l_{4}}^{l_{1},l_{2}}\, \hat{a}^{\dag}_{l_{1}}\hat{a}^{\dag}_{l_{2}}
\hat{a}_{l_{3}}\hat{a}_{l_{4}},
\end{equation}
where $l_i = 3m$ and $m=0,1,\ldots$. This Hamiltonian can be realized by coupling the cavity field $\hat{\mathcal{E}}_{0,0}$ to a high-lying Rydberg state in an EIT configuration \cite{Pritchard2010,Gorshkov2011,Petrosyan2011}. In recent cavity experiments the strong nonlinearity on the single photon level was demonstrated \cite{Ningyuan2016,Ningyuan2017}. The Rydberg cavity polaritons have an effective interaction potential $V(r)=C_{6}/(r^{6}+a_{B}^{6})$. Here, $C_{6}$ is the effective interaction strength and $a_{B}$ is the Rydberg blockade radius. Although the opposite regime is very interesting on its own right \cite{Grusdt2013}, we assume in the following the case where the magnetic length $l_{B}=w_{0}/2$ is much larger than $a_{B}$. In this limit the dominant interaction contribution comes from the zero's Haldane pseudo potential  \cite{Grusdt2013}
\begin{equation}
V_0 \simeq \frac{3C_6}{8 l_B^2 a_B^4},
\end{equation}
 which determines all interaction coefficients \cite{Jain2007}
\begin{align}
V_{l_{3},l_{4}}^{l_{1},l_{2}} &= \langle l_1,l_2 | \hat{V} | l_3, l_4 \rangle \nonumber \\
&\simeq \frac{V_0}{2}(l_1+l_2)! \sqrt{\frac{2^{-2(l_1+l_2)}}{l_1!l_2!l_3!l_4!}} \delta_{l_1+l_2,l_3+l_4}.
\end{align}
We assume for the interaction coefficients $V_{l_{3},l_{4}}^{l_{1},l_{2}} \ll \vert\Delta_{0,1}-\Delta_{0,0}\vert$, i.e. they are small compared to the energy gap between the Landau levels to avoid mixing of states in different Landau level.

The combination of the photonic Landau Level Eq. \eqref{eq:PhotonicLandauLevel} and the strong photon nonlinearity Eq. \eqref{eq:PhotonPhotonInteraction} lead to a set of degenerate low energy states with total angular momentum $L$ depending on the photon number $N$ \cite{Sommer2015,Schine2016,Wu2017}. For a given photon number $N$ the zero energy state with lowest total angular momentum,
\begin{equation}
\langle z_1,\ldots,z_N|\mathrm{LN},N \rangle = \prod_{i<j}(z^{3}_{i}-z^{3}_{j})^{2},\label{eq:LNState}
\end{equation}
is a unique ground state of the system \cite{Wu2017}, which resembles a Laughlin-type state. We here have dropped the ubiquitous Gaussian factor and the normalization constant \cite{Schine2016,Wu2017}. The two-dimensional coordinate is $z_{j}=x_{j}-i y_{j}$. The total angular momentum of the state \eqref{eq:LNState} with $N$ photons is $\hat{L}|{\rm LN},N\rangle=3N(N-1)|{\rm LN},N\rangle$. In addition, we consider here the $m$th quasi-hole states with $N$ photons
\begin{equation}
\langle z_1,\ldots,z_N|{m{\rm qh}} \rangle=\prod_{k}z_{k}^{3m}\prod_{i<j}(z^{3}_{i}-z^{3}_{j})^{2},
\label{eq:mQuasiHoleState}
\end{equation}
having total angular momentum $\hat{L}|m{\rm qh},N\rangle=\frac{3}{2}mN(N+1)|m{\rm qh},N\rangle$. It is straightforward to show that the Laughlin-type state with $N+1$ photons has the same total angular momentum as the 2-quasi-hole state with $N$ photons.

\subsection{Full Protocol}

\paragraph*{Single Photon Pump. --}

\begin{figure}
\includegraphics[width=0.45\textwidth]{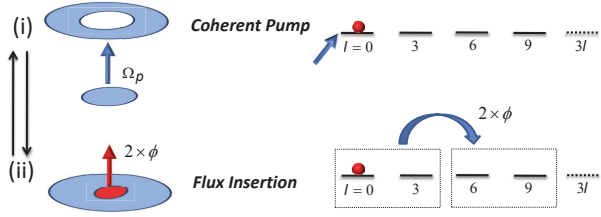}
\caption{(Color online) The growing scheme which is used for the preparation of the Laughlin-type states consists of two steps. (i) Coherent pump of a single photon in the ground state of the cavity by using the non-linear photon-photon interaction. (ii) Increase of angular momentum per particle by $6$ (flux insertion). Repeating the two steps to grow a photon Laughlin-type state.}
\label{LNf}
\end{figure}

We consider a coherent pump which injects a single photon into the mode $\hat{a}_{0}$. This implies that there is no transfer of angular momentum into the system. We assume that an external laser field is applied with mode profile matching the $l=0$ angular momentum state,
\begin{equation}
\hat{H}_{\Omega_{p}}=\Omega_{p}(\hat{a}_{0}^{\dag}e^{-i\omega t}+\hat{a}_{0}e^{i\omega t}).\label{H_Omega}
\end{equation}
Here $\Omega_{p}$ is the driving pump Rabi frequency into the cavity and $\omega$ is the oscillation frequency which we assume to be in resonance with respect to the energy of the LLL, i.e. $\omega = \Delta_{0,0}$. Without the interaction Eq. \eqref{eq:PhotonPhotonInteraction} the Hamiltonian (\ref{H_Omega}) creates a coherent amplitude of the photonic mode which contains a superposition of many photons. However,  strong photon blockade ensures the insertion of a single photon requiring $\Omega_{p} \ll V_0,\Delta_\mathrm{LN}$, where $\Delta_\mathrm{LN} \simeq 0.2 V_0$ is the many body gap of the system. Note that the Laughlin gap only slightly depends on the photon number $N$. Starting from a 2 quasi-hole state $|2{\rm qh},N\rangle$, we use a $\pi$-pulse of time $\tau_p = \pi/2\Omega^{(N)}$ where
\begin{equation}
	\Omega^{(N)}= \Omega_p \langle {\rm LN},N+1|\hat{a}_{0}^{\dag}|2{\rm qh},N\rangle
\end{equation}
is the coupling between the quasi-hole and Laughlin state \cite{Grusdt2014,Letscher2015}, to insert a single photon.

\paragraph*{Adiabatic Flux Insertion.--}

In Sec. \ref{FInsertion} we discussed the noninteracting case for inserting flux quanta by transferring all photons to the first Landau level and then back to the LLL. Now, in the interacting case we require adiabaticity $\Delta_\mathrm{LN}\tau_\f \gg 1$, where the many-body gap $\Delta_\mathrm{LN}$ should not vanish during flux insertion. To this end we couple the photonic cavity field $\hat{\mathcal{E}}_{0,1}$ in the first Landau level using an EIT scheme to a Rydberg state as well. This ensures to maintain a finite many-body gap $\Delta_\mathrm{LN}$. For simplicity we assume the same interaction potential $V(r)$ as before. Now, in Eq. \eqref{eq:PhotonPhotonInteraction} we sum over all photonic modes in the lowest and first Landau level.

\paragraph*{Protocol.--}

The growing scheme is depicted in Fig. \ref{LNf}. It starts by preparing the cavity with no photon. Then in the first step a single photon in mode $\hat a_{0}$ is pumped into the cavity $|0\rangle\rightarrow \hat{a}_{0}^{\dag}|0\rangle$ by using the non-linear interaction Eq. \eqref{eq:PhotonPhotonInteraction}. This state obviously has total angular momentum $L=0$. Next we repeat the flux insertion scheme in Sec. \ref{FInsertion} two times which realizes the transition $\hat{a}^{\dag}_{0}|0\rangle\rightarrow \hat{a}^{\dag}_{6}|0\rangle$ with $L=6$. The latter state is a 2 quasi-hole state with one photon. Now a second photon is pumped into the cavity. The finite overlap $\Omega^{(1)}/\Omega_p = \sqrt{10/11}$
with the Laughlin state ensures that we pump into the ground state of the system. This step creates a Laughlin state with $N=2$ photons. By repeating these two steps we grow a Laughlin-type state \eqref{eq:LNState} with $N$ photons.

\begin{figure}
\includegraphics[width=0.45\textwidth]{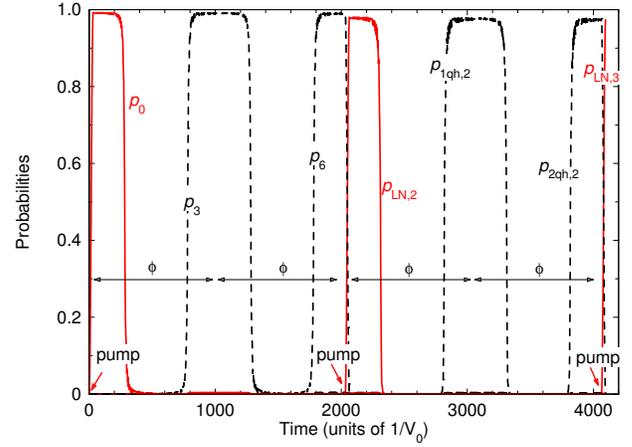}
\caption{(Color online) Numerical simulation of growing scheme for creation of Laughlin states with $N=2$ and $N=3$ photons. The system is prepared initially in a state with no photon. The red arrow indicates the time at which the coherent pump is applied. The angular momentum per photon is increased by repeating twice the flux insertion. The probabilities are $p_{m}=|\langle \psi|a_{m}^{\dag}|0\rangle|^{2}$, and respectively $p_{{\rm LN},N}$ and $p_{m{\rm qh},N}$ are the probabilities for Laughlin state and $m$th quasi-hole state. The adiabatic flux insertion method for numerical simulation is explained in Appendix \ref{ap:FluxInsertionMethod2}. The parameters are set to $\Delta_{0}/V_{0}=10$, $\Omega_{p}/V_{0}=1/20$, $g_a/V_{0}=g_b/V_{0}=1/5$.  }
\label{LN_N}
\end{figure}

To numerically simulate the full growing protocol is rather involved, since taking into account all different atomic excitations leads to fast growing of the relevant Hilbert space even for few excitations. Therefore we simplify the protocol reducing it to the essential components, namely the adiabatic increase of angular momentum of the cavity modes by flux insertion and subsequent photon insertion. The simplified
flux insertion method used for the simulation relies solely on the photonic cavity modes and is therefore amenable to numerical simulations by
exact diagonalization.  Specifically we consider a direct coupling between the lowest and first Landau level and change the energies of the Landau levels in time. This resembles a rapid adiabatic passage sweep. In Appendix \ref{ap:FluxInsertionMethod2} we explain the method in detail. In Fig. \ref{LN_N} we show a numerical simulation of the full protocol for the preparation of Laughlin states up to three photons.  After three steps of the protocol we obtain a LN state with three photons with probability $|\langle\psi|\rm{LN},3\rangle|^{2}\approx 0.97$.

Finally, let us comment on the fidelity of our scheme. On the one hand, in the flux insertion process, see Sec. \ref{Imperfections}, the imperfections come from non adiabatic transitions, which requires $\Omega_l\tau_\f \gg 1, \Delta_\mathrm{LN}\tau_\f \gg 1$. On the other hand, in the coherent pump the imperfections come from coupling to higher photon number states which require $\Delta_\mathrm{LN}\tau_p \gg 1$. While both favor large timescales $\tau = 2\tau_\f +  \tau_p$ for each step in the growing protocol, losses limit the timescale $\tau$.
We take into account the effect of cavity losses as well as the finite lifetime of the Rydberg state by an effective
 loss rate $\gamma_\mathrm{eff}$.  As shown in Ref. \cite{Grusdt2014,Letscher2015}, the fidelity for the creation of an $N$-photon Laughlin state then scales as
\begin{equation}
\mathcal{F}_N \simeq \exp \left[ - \frac{1}{2}N \left(
\frac{1}{2} \gamma_\mathrm{eff} \tau (N+1) + \frac{\Lambda_N^2}{(\Delta_\mathrm{LN}\tau)^2}
 \right)\right],
\end{equation}
where $\Lambda_N$ depends on photon number $N$. Note that our protocol first creates a hole excitation in the center and then refills the hole. Repeating the steps of the protocol  photons are pumped continuously into the center of the system. Defects created by losses will
be continuously pumped to the periphery of the system and we expect that a much higher fidelity can be achieved in the steady state in the center of the cavity.

\section{Discussion and Outlook}\label{C}

In summary, we discussed an adiabatic transfer protocol to insert flux quanta in a photonic twisted cavity setup. The scheme relies on a robust STIRAP technique transferring OAM of an external classical laser beam to the photonic cavity modes. A dense atomic ensemble hereby acts as a mediator. We show that the transfer can be described by a set of dark state polaritons between cavity modes with different angular momentum. Furthermore, we discuss imperfections of the protocol and estimate the fidelity. In addition we discuss the preparation of Laughlin-type states based on the growing protocol of Refs.\cite{Grusdt2014, Letscher2015}. To this end, we discuss a single photon pump coupling the cavity field to a high-lying Rydberg state in an EIT configuration. We show that by successive repetition of flux insertion and coherent pump a Laughlin-type state can be prepared with high fidelity. Since as compared to alternative growing protocols \cite{Dutta2017} in our scheme photons and thus also loss-induced defects are continuously pumped from the center to the periphery of the system, we expect to create Laughlin-type states with much higher fidelity in the center of the cavity.

The non-local character of the interaction between Rydberg polaritons may lead to
other interesting states such as the Moore-Read Pfaffian \cite{Moore1991} in the regime of large magnetic fields, where the magnetic
length becomes comparable or smaller than the blockade radius \cite{Grusdt2013}.  Furthermore, the coherent control may allow to investigate bilayer quantum Hall phases exploring different photonic Landau levels.

\section*{Acknowledgments}
This work was supported by the DFG through the priority program GiRyd (SPP 1929). PAI acknowledges support by the ERyQSenS, Bulgarian Science Fund Grant No. DO02/3. The authors thank D. Dzsotjan for helpful discussions.

\appendix
\section{Coupling Matrix Elements }\label{LGP}

\subsection{Properties of Laguerre-Gauss Polynomials}

The Laguerre-Gauss (LG) polynomials form a complete set of functions with the following orthogonal condition
\begin{equation}
\int_{0}^{\infty}e^{-x}x^{k}L_{n}^{k}(x)L_{m}^{k}(x)dx=\frac{(n+k)!}{n!}\delta_{n,m}.\label{orth}
\end{equation}
For $n=0$ we have $L_{0}^{l}(x)=1$. The LG polynomials also obey the recurrence relations $L_{n}^{l-1}(x)=L_{n}^{l}(x)-L_{n-1}^{l}(x)$ and $\sum_{p=0}^{n}L_{p}^{l}(x)=L_{n}^{l+1}(x)$.

\subsection{First Step}

The coupling matrix elements of the first step are
\begin{equation}
\chi_{3m}^{n^{\prime},n}
=C_{3m}^{n^{\prime},n}\int_{0}^{\infty}e^{-2x^{2}}x^{6m+2}\kappa(x)L_{n}^{3m+1}(2x^{2})L_{n^{\prime}}^{3m}(2x^{2})dx,\label{I}
\end{equation}
with $C_{3m}^{n^{\prime},n}=2^{3m+2}\sqrt{\frac{2n!n^{\prime}!}{(3m+1+n)!(3m+n^{\prime})!}}$. Here the function $\kappa(x)$ describes the
intensity shape of the Rabi frequency $\bar{\Omega}_{1}$. We choose
\begin{equation}
\kappa(x)=\frac{x^{2}}{a^{3}+x^{3}},
\end{equation}
where $a=r_{0}/w_{0}$ is the dimensionless cutoff.

\paragraph*{$a\to 0$ limit:}

Considering first the LLL, i.e. $n=0$ and making the substitution $y=2x^{2}$ we have
\begin{equation}
\chi_{3m}^{n^{\prime},0}=\sqrt{\frac{2n^{\prime}!}{(3m+1)!(3m+n^{\prime})!}}\int_{0}^{\infty}e^{-y}y^{3m}L_{n^{\prime}}^{3m}(y)dy.
\end{equation}
Using the orthogonality Eq. (\ref{orth}) we find
\begin{equation}
\chi_{3m}^{n^{\prime},0}=\sqrt{\frac{2}{3m+1}}\delta_{0,n^{\prime}}.
\end{equation}
The general result for the coupling coefficient in the limit $a\rightarrow 0$ is
\begin{displaymath}
\chi_{3m}^{n^{\prime},n} = \left\{
\begin{array}{ll}
 \sqrt{\frac{2(3m+n^{\prime})!n!}{n^{\prime}!(3m+1+n)!}} & \quad\textrm{for}\quad n\geq n^\prime,\\
 \phantom{\Bigl(}0 & \quad\textrm{for}\quad n< n^\prime.
 \end{array}\right.
\end{displaymath}
While the couplings $\chi_{3m}^{0,n^\prime}$ between the lowest and higher Landau levels  $n^\prime \geq 1$
are of order unity even for $a\rightarrow 0$, all couplings $\chi_{3m}^{n^\prime,0}$ vanish identically.

\paragraph*{lowest order corrections in $a$:}

The lowest order correction of $\chi_{3m}^{n^\prime,0}$ in $a$ for $m=0$ is
\begin{equation}
\chi_{0}^{n^{\prime},0}\approx -\frac{8\pi}{3}\sqrt{\frac{2}{3}}a^{2}+{\cal O}(a^{3})
\end{equation}
and becomes even smaller for $m>0$. Thus as long as $a\ll 1$ coupling to higher Landau levels is negligible.

\subsection{Second Step}

In the second step we choose
%
$\bar{\Omega}_{2}(r,\varphi,t)=\Omega_{2}(t)\, x^{2}\, e^{2i\varphi},$
%
such that the coupling strengths become
\begin{equation}
\tilde{\chi}_{3m}^{n^{\prime},n}=\tilde{C}_{3m}^{n^{\prime},n}
\int_{0}^{\infty}e^{-2x^{2}}x^{6m+1}L_{n}^{3m-2}(2x^{2})L_{n^{\prime}}^{3m}(2x^{2})dx,
\end{equation}
with $\tilde{C}_{3m}^{n^{\prime},n}=2^{3m+1}\sqrt{\frac{n!n^{\prime}!}{(3m-2+n)!(3m+n^{\prime})}}$.
Setting $n=0$ and making the substitution $y=2x^{2}$ we get
\begin{eqnarray}
\tilde{\chi}_{3m}^{n^{\prime},0}=\frac{1}{2}\sqrt{\frac{n^{\prime}!}{(3m-2)!(3m+n^{\prime})!}}\int_{0}^{\infty}e^{-y}y^{3m}L_{n^{\prime}}^{3m}(y)dy.
\end{eqnarray}
Finally, using the orthogonality Eq. (\ref{orth}) we obtain
\begin{eqnarray}
\tilde{\chi}_{3m}^{n^{\prime},0}=\frac{1}{2}\sqrt{\frac{3m!}{(3m-2)!}}\delta_{0,n^{\prime}},
\end{eqnarray}
which implies that the transition to states with $n^{\prime}>0$ are completely suppressed.


\section{Non-Adiabatic Losses}\label{NAL}

Consider the five-level system driven by two cavity fields $g_{3m}$ and $g_{3m+1}$ and respectively two classical laser beams $\Omega_{1}$ and $\Omega_{0}$ as is depicted in Fig. \ref{fig22}. Including the spontaneous decay from the two excites states the non-hermitian interaction Hamiltonian becomes
\begin{equation}
H =\left[\begin{array}{ccccc}
-i\gamma & \Omega_{1}&0&g&0  \\ \Omega_{1}
& 0&\Omega_{0}&0&0 \\ 0&\Omega_{0}&-i\gamma&0&g\\g&0&0&0&0\\0&0&g&0&0
\end{array}\right],\label{F}
\end{equation}
where for simplicity we assume equal cavity couplings $g_{3m}=g_{3m+1}=g$. The time-dependent Schr\"odinger equation reads $i\hbar\frac{d}{dt}\vec{B}=H\vec{B}$ with column vector $\vec{B}=[c_{e},c_{s},c_{r},c_{a_{1}},c_{a_{2}}]^{\rm T}$ comprises the diabatic probability amplitudes of the five states. For $\gamma=0$ the eigen-spectrum of $H$ consists of one zero energy dark state
\begin{equation}
|d\rangle=\cos(\varphi)\sin(\theta)|a_{0}\rangle+\sin(\varphi)\sin(\theta)|a_{1}\rangle-\cos(\theta)|s\rangle,
\end{equation}
two states with energies $E_{\pm}=\pm g$,
\begin{eqnarray}
|\pm g\rangle=\frac{1}{\sqrt{2}}\Bigl(\sin(\varphi)|e\rangle-\cos(\varphi)|r\rangle\pm\sin(\varphi)|a_{0}\rangle\mp\cos(\varphi)|a_{1}\rangle\Bigr)
\end{eqnarray}
and two states with energies $E_{\pm\lambda}=\pm\lambda$ with $\lambda=\sqrt{g^{2}+\Omega^{2}_{1}+\Omega^{2}_{0}}$,
\begin{eqnarray}
|\pm\lambda\rangle&=&\frac{1}{\sqrt{2}}(\pm\cos(\varphi)|e\rangle+\sin(\theta)|s\rangle\pm\sin(\varphi)|r\rangle\notag\\
&&+\cos(\varphi)\cos(\theta)|a_{0}\rangle+\sin(\varphi)\cos(\theta)|a_{1}\rangle)
\end{eqnarray}
Here the states $\vert a_j\rangle \equiv \vert g\rangle \hat a_j^\dagger \vert 0\rangle$ denote the states with one photon in mode $\hat a_j$
and all atoms in the ground state.
The mixing angles are defined by
\begin{equation}
\tan(\varphi)=\frac{\Omega_{0}}{\Omega_{1}},\quad\tan(\theta)=\frac{\sqrt{\Omega^{2}_{1}+\Omega^{2}_{0}}}{g}.
\end{equation}
Note that due to the spatial dependence of the laser fields, the mixing angles vary with the distance to the origin which could violate the adiabaticity of the transition close to the cavity axis.

It is convenient to work in the adiabatic basis, where the loss of transfer efficiency due to the spontaneous decay shows up as population decay from the dark state. The probability amplitudes $\vec{A}=[a_{g},a_{-g},a_{d},a_{-\lambda},a_{\lambda}]^{\rm T}$ of the adiabatic states are connected to the diabatic amplitudes by the relation $\vec{A}=W\vec{B}$, where the orthogonal rotation matrix $W$ is given by

\begin{widetext}
\begin{equation}
W =\frac{1}{\sqrt{2}}\left[\begin{array}{ccccc}
\sin(\varphi) & 0&-\cos(\varphi)&\sin(\varphi)&-\cos(\varphi)  \\ \sin(\varphi)
& 0&-\cos(\varphi)&-\sin(\varphi)&\cos(\varphi) \\ 0&-\sqrt{2}\cos(\theta)&0&\sqrt{2}\cos(\varphi)\sin(\theta)&\sqrt{2}\sin(\varphi)\sin(\theta)\\-\cos(\varphi)&\sin(\theta)&-\sin(\varphi)&\cos(\varphi)\cos(\theta)
&\sin(\varphi)\cos(\theta)\\\cos(\varphi)&\sin(\theta)&\sin(\varphi)&\cos(\varphi)\cos(\theta)&\sin(\varphi)\cos(\theta)
\end{array}\right].\label{W}
\end{equation}
\end{widetext}

It is straightforward to show that the Schr\"odinger equation in the adiabatic basis reads $i\hbar\frac{d}{dt}\vec{A}=W^{-1}HW-iW\frac{d}{dt}W^{-1}\vec{A}$, where the last term describes the non-adiabatic transitions.
As long as the spontaneous decay is sufficiently strong the populations of the non-zero energy adiabatic states change negligibly such that one can perform adiabatic elimination $\dot{c}_{\pm g}=\dot{c}_{\pm\lambda}=0$. Assuming $g\gg \frac{\dot{\varphi}^{2}}{\lambda}\cos(\theta)$ and neglecting terms of order of $O(\dot{\varphi}^{3})$  $O(\dot{\theta}^{3})$, $O(\dot{\varphi}^{2}\dot{\theta})$ and $O(\dot{\varphi}\dot{\theta}^{2})$ we find
\begin{equation}
P_{\rm d}\approx\exp\biggl\{-\frac{2\gamma}{g^{2}}\int_{-\infty}^{\infty}\Bigl(\dot{\varphi}^{2}\sin^{2}(\theta)+
\dot{\theta}^{2}\cos^{2}(\theta)\Bigr)dt\biggr\}.
\end{equation}
One can evaluate the probability remaining in the adiabatic states during the both stages of the protocol using the Rabi frequnecies (\ref{Rabi}) and (\ref{Rabi2}). We obtain $P_{\rm d}(\vec{r})=e_{a}(\vec{r})$ with $a=1,2$, where
\begin{equation}
e_{a}(\vec{r})=\exp\biggl\{-\frac{\gamma}{4T}\left(\frac{2}{g^{2}}-\frac{1}{g^{2}+\Omega^{2}}-\frac{1}{g^{2}+f^{2}_{a}\Omega^{2}}\right)\biggr\}.
\end{equation}
Here $f_{1}=\kappa(x)$ and $f_{2}=x^{2}$.

\section{Flux insertion method for numerical simulation}
\label{ap:FluxInsertionMethod2}

Simulating the flux insertion protocol in Sec. \ref{FInsertion} in the presence of particle-particle interactions is numerically challenging since it requires to include all photonic modes as well as a large number of four-level atoms. In order to simplify the numerical integration we replace the flux insertion technique based on STIRAP by another adiabatic transfer technique without coupling to an atomic medium. This protocol cannot directlty be realized with the current cavity setup, but correctly captures the fidelity of the growing protocol for Laughlin states.

As before, we split the protocol into two steps,
\begin{equation}
{\rm i)}:\quad\hat{a}_{0,3m}\rightarrow \hat{a}_{0,3m+1},\qquad{\rm ii}): \quad \hat{a}_{0,3m+1}\rightarrow \hat{a}_{0,3m+3}.
\end{equation}
To mimic the adiabatic transfer mediated by the atomic medium in an effective way, we add to the photonic Landau level Hamiltonian, Eq. \eqref{eq:PhotonicLandauLevel}, a weak coupling between the lowest and first Landau level with time-dependent couplings
\begin{equation}
\hat{H}_{c}= \sum_{m} \left[ g_{a}(t) \hat{a}_{3m}^{\dag}\hat{a}_{3m+1} + g_{b}(t) \hat{a}_{3m+1}^{\dag}\hat{a}_{3(m+1)}+{\rm h.c.} \right]
\end{equation}
The coupling $g_{a}(t)$ is on during step (i) and $g_{b}(t)$ during step (ii). Furthermore $g_a,g_b \ll \Delta_\mathrm{LN}$. We assume that the excited state Landau level energy can be changed linear in time $t$ with respect to the lowest Landau level energy,
\begin{equation}
\Delta(t) = - \Delta_0+\frac{4\Delta_0}{\tau_\f}\left|t-\frac{\tau_\f}{2}\right|,
\end{equation}
where $\Delta_0 = \Delta_{0,1} - \Delta_{0,0}$. This resembles a rapid adiabatic passage protocol. Let us start with the adiabatic limit $t_\f \rightarrow \infty$.
In the first step, the coupling $g_a$ is turned on and we transfer an OAM per photon of $1\hbar$ after the first step $t = t_\f/2$. Then, $g_a$ is turned off and $g_b$ is turned on. The second step starts with detuning $\Delta(t_\f/2) = -\Delta_0$ and transfers $2\hbar$ flux per photon to the system after time $t=t_\f$. After a full step we added $3$ flux quanta per particle to the system. For adiabaticity, we require
\begin{equation}
\tau_\f \gg \frac{4\Delta_0}{\Delta_\mathrm{LN}^2},
\end{equation}
which means that the detuning sweep $\Delta(t)$ must be slow compared to the many-body gap of the system.

\end{document}